\documentclass[12pt, draftclsnofoot, onecolumn]{IEEEtran}
\usepackage{wrapfig}
\usepackage{color}
\usepackage{array}
\usepackage{float}
\usepackage{graphicx}
\usepackage{enumitem}
\newcolumntype{P}[1]{>{\centering\arraybackslash}p{#1}}
\usepackage{hyperref}
\usepackage[export]{adjustbox}
\usepackage{multirow}
\usepackage{hyperref}
\usepackage[export]{adjustbox}
\usepackage{multirow}
\usepackage{tikz}
\usepackage{caption}
\usepackage{subcaption}
\usepackage{amsmath}
\usepackage{ulem}

\begin{document}

\author{\IEEEauthorblockN{Santosh Ganji, P. R. Kumar, Fellow, IEEE}\\
\IEEEauthorblockA{Texas A\&M  University}}

\title{Seeing the Unseen: The REVEAL protocol to expose the wireless Man-in-the-Middle}

 \newcommand{\prk}[1]{{\color{blue}\textit{Professor: #1}}}
 \newcommand{\sg}[1]{{\color{red}\textit{Santosh: #1}}}

 \maketitle

\begin{abstract}
A Man-in-the-Middle (MiM) can collect over-the-air packets whether from a mobile or a base station, process them, possibly modify them, and forward them to the intended receiver. This paper exhibits the REVEAL protocol that can detect a MiM, whether it has half-duplex capability, full-duplex capability, or double full-duplex capability.
The REVEAL protocol creates a sequence of timing-based challenge packets where the transmission times of the packets, their durations, and their frequencies, are chosen to create conflicts at the MiM, and make it impossible for the MiM to function.
Implementing the REVEAL protocol in 4G/5G technology, we instantiate a MiM between the 4G/5G base station and a mobile, and exhibit the successful detection mechanisms. 
With the shared source code, the MiM can be reproduced using software defined radios and protocol efficacy can be verified using any open software defined cellular networks with off-the-shelf devices.
\end{abstract}

\section{Introduction}
 A Man-in-the-Middle (MiM) collects over-the-air radio samples, whether transmitted by a mobile or the base station, processes the message, possibly modifies the contents of the packet \footnote{In this work, we assume the MiM cannot decode encrypted traffic.}, and forwards the packet.
The two endpoints of the thus created link may be oblivious to the presence of the MiM.
The challenge addressed in this paper is how
to detect the presence of a MiM that wants to stay hidden.

\begin{figure*}[h]
    \centering
    \includegraphics[width=.9\linewidth]{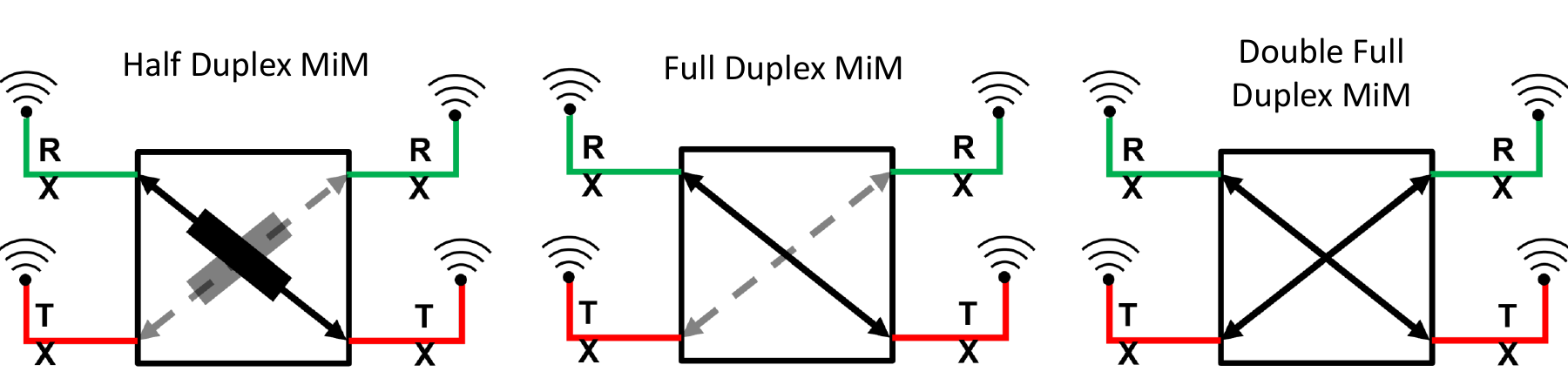}
    \caption{Three classes of Man-in-the-Middle}
    \label{MiMs}
\end{figure*}

An adversary follows a procedure, also called attack vector, to orchestrate an MIM between the base station and the mobile. At the radio level, irrespective  of the attack vector, 
the MiM can have one of three packet forwarding capabilities - Half duplex, Full-duplex or Double full-duplex, as shown in Fig.~\ref{MiMs}. A half duplex node cannot transmit while listening, a full-duplex node can forward messages while listening but only in one direction, and a double full-duplex node can simultaneously listen and forward messages in both directions.\\
Broadly speaking, vulnerability researchers discover new attack vectors to infiltrate networks, while security researchers work to close the loopholes in the system to prevent these attack vectors from functioning. In the field of wireless networks, particularly cellular networks, several vulnerabilities have been identified for launching a MiM attack, and solutions have been proposed to mitigate these attacks. In Section \ref{related_work}, we present MiM attacks that have been discovered and addressed across generations of cellular networks. To detect an ongoing attack, previously proposed ideas rely on either the signature of the attack vector or the detection of anomalous behavior. However, this makes detection challenging, as nodes in the network must monitor several indicators, and such previous approaches cannot detect newly discovered attack vectors, unlike our approach. 

We introduce a protocol, REVEAL, which can detect a MiM with any one of the above three forwarding capabilities and does not rely on any particular attack vector or the monitoring of anomalous behavior of nodes.

This protocol is based on the initial theoretical work in \cite{MiM_Kumar} for 
Half-duplex and Full-duplex MiMs.
The REVEAL protocol is designed to function so that it is agnostic to the attack vector. 
We experimentally evaluate the REVEAL protocol. We also show the efficacy of the solution and relevance for the current day 4G/5G cellular wireless technology.

A MiM  
adds  finite delay to the wireless traffic.
Consider a simple MiM-Relay. 
The relay either immediately re-transmits the incoming signal, or demodulates and re-modulates the packet before forwarding the received packet. The works  \cite{relay_delay_analysis,relay_capacity} show that even a simple MiM acting like a relay that processes the message only at the radio layer delays the traffic. A more sophisticated MiM  may perform several data processing tasks \cite{adaptover_mim1,mitm1}. 
The MiM attack \cite{mitm1} that claims to break 4G communication networks processes data traffic all the way up to the data link layer, the top most layer of the 4G/5G protocol stack. Also, after processing the packet at the desired layer in the protocol state, the MiM reconstructs the over-the-air packet, all the way down to the physical layer. This entails a delay before the packet is relaunched into the wireless medium.

In a communication network, wired or wireless,
all nodes must maintain a common time despite having independent 
clocks. 
Clock synchronization is necessary to establish a common time in a communication network. For example, a base station and a mobile must have a common time to schedule communication opportunities.
Any communication technology, wired or wireless, has an associated clock synchronization mechanism. 
In particular for cellular networks, where the location of a mobile changes as the user moves, a significant portion of protocol design is dedicated for clock synchronization \cite{4G_sync0,3gpp_specifications}.

A natural question one can ask is: Can we devise  mechanisms taking advantage of clock synchronization to detect a MiM? This question was first examined in \cite{MiM_Kumar}, where the cases of half duplex and full-duplex MiMs were investigated. In this paper, we examine the detection mechanisms for all the three types of MiM nodes. We show how to detect double full-duplex MiMs, and present experimental results for all three scenarios. We  show that the entire end-to-end scheme including clock synchronization can be realized over software defined radios, and present the performance results. To demonstrate them in action, we built MiM nodes using software defined radios. Using the software-defined radio MiM we launch attacks on the 4G and 5G networks. We show that the mobile naively connects to the cellular network totally ignorant of the MiM. Then we use our mechanism to catch the MiM nodes. Thereby we show that it is indeed possible to detect middle nodes, irrespective of the signature of the attack or attack vector of MiMs.
The design and source code of the software defined MiM is available at \cite{MiM_Full_Duplex_Implementation}. 

\begin{figure*}[t!]

\begin{minipage}[b]{.5\textwidth}
\includegraphics[width=.8\linewidth]{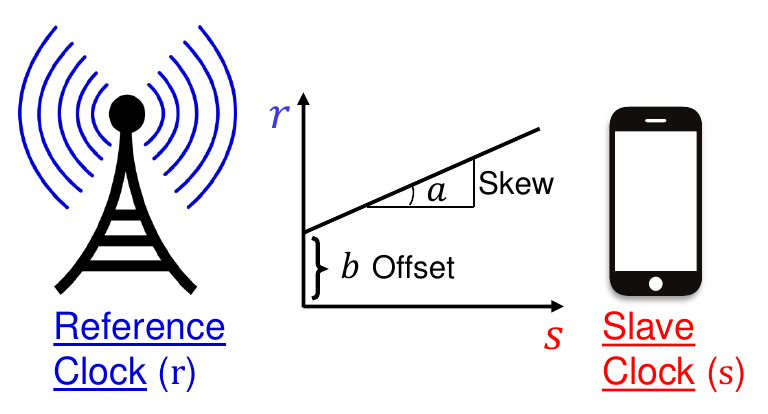}
\caption{Relationship between reference and slave clocks} \label{fig:clock_relationship}
\end{minipage}
\hspace{.01cm}
\begin{minipage}[b]{.5\textwidth}
\includegraphics[width=.8\linewidth]{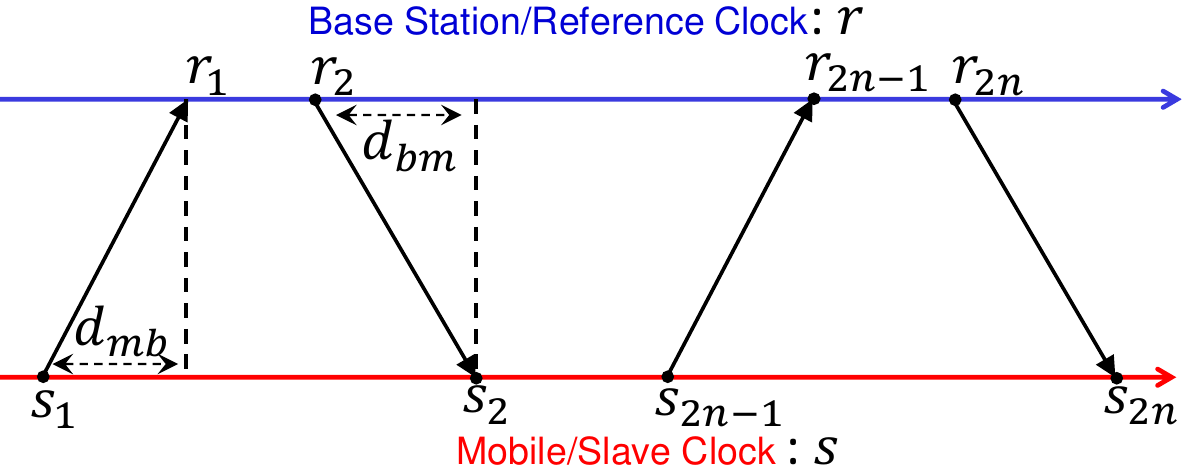}
\caption{Timestamp exchanges. (Note that $d_{mb}$ is to be measured in the reference clock's time units rather than the slave clock's.)  } \label{fig:time_exchanges}
\end{minipage}


\end{figure*}

Summary of contributions in the work:

\begin{itemize}
        \item One of the challenges we address to detect an MiM is to show the feasibility of an MiM attack on cellular networks: We built half, full and double full-duplex MiM nodes.
    \item  With our implementation, we show a live MiM attack on 4G network: both full-duplex MiM attack and double full-duplex MiM attack.

    \item We present methods for detecting a Man-in-the-Middle (MiM) with any of these three capabilities. Detecting a Half-duplex MiM involves sending long packets, identifying a Full-duplex MiM entails scheduling bi-directional traffic, and recognizing a double full-duplex MiM requires scheduling swift handovers.
 
    \item We devised a usable protocol that detects presence of a MiM.
    \item We demonstrate the working of our protocol on standard compliant 4G Network.
       \item Our MiM hardware and software designs are available to reproduce the attack and detection methods.
    \item 
    Overall, this work is the first practical implementation of MiM detection methods on 4G/5G networks.

\end{itemize}

The rest of the paper is organized as follows. In Section \ref{background}, we address clock synchronization, and present how a MiM node impacts the clock synchronization. We present our detection mechanisms in Section \ref{detection}.
The MiM node implementation details, and attack and detection on a 4G/5G network, are described in Section \ref{implementation}.    
\section{Related Work}\label{related_work}

We first review several MiM attack vectors that target commonly employed wireless technologies. Elaborating on how a MiM forwards the messages in the link, we present several proposals from the literature to detect the MiM. 

The key differentiating factor of our REVEAL protocol is that it catches any middle node, whether it is half, full or double full-duplex. To the best of our knowledge REVEAL is the only work that condenses the detection mechanisms into three fundamental principles that do not require any additional information on the attack vector, technology or fingerprints whatsoever. Therefore REVEAL can continue to detect future MiM attacks as communication protocols and technology continue to progress.
\subsection*{IEEE 802.11 Family}
Hwang et al. \cite{wifi1} have shown how to attack 802.11a/b/g networks using a rogue access point capitalizing on  the security vulnerabilities of the authentication framework. Lynn and Baird have presented a "Monkey Jack" attack that inserts a MiM between an 802.11b station and an access point \cite{lynn_baird}. The attacker listens to the channel of the access point, but forwards on a different channel. Ignorant of this fact, the wifi station connects to the attacker. To mitigate such attacks, the authors have proposed strong authentication and radio signal shaping such that the wireless nodes transmit in a directional fashion and with low power. Serious flaws in the Wired Equivalent Privacy of 802.11b make it possible for a MiM to even modify the contents of the packets \cite{cam2003security}. To mitigate this, improved encryption mechanisms have been proposed, but the superseding encryption mechanisms of Wifi Protected Access (WPA) and WPA-2 are also found to be vulnerable \cite{bradbury2011hacking}. Works \cite{ahmad2010wpa,herzberg2009stealth,7031876} detail the MiM attacks. Kumar et al. \cite{6449834} have proposed modifications to the vulnerable protocol to mitigate the attack. Recently, Steinmetzer et al. \cite{beam_stealing} has devised a way to launch an MiM attack on a 802.11ad mm-wave network, and observed that anomalies in received signal strength and modifying beacon configurations help detection.

At the radio level, all the attacks need either a half or full-duplex MiM. 
Our full-duplex detection method detects the presence of all such MiMs. 
Given the constant discovery of new exploits across the generations of wireless standards, our attack agnostic method serves as a useful tool to detect MiMs.
\\ 
\subsection*{Cellular Technology}
\subsubsection*{2G Technology} Strobel \cite{strobel2007imsi} has discussed an attack on a 2G network where the attacking device is a half duplex MiM that alternately impersonates  the base station and the mobile, and taps the communication link.
\noindent
\subsubsection*{3G Technology} The authors of \cite{2G_3G_Mid_generation} have found a vulnerability to downgrade a mobile connected to a 3G network to 2G, after which a MiM attack on 2G is launched. Due to the lack of message integrity check, the MiM eavesdrops on the mobile's communication with the 2G base station. 
Beekman et al. discovered an attack vector that exploits weak authentication to launch an MiM attack that can modify the routes and contents of T-Mobile's WiFi calling service \cite{wifi_calling_tmobile}. Improved verification brought closure to the attack. 
\noindent
\subsubsection*{4G/5G Technology} The authors of \cite{lte_security_disabled,4G_5G,mitm1} found missing integrity checks in some network configurations of 4G and early 5G protocols, and demonstrated the feasibility of a MiM. In these attacks,  the MiM relays modified information between the mobile and the base station. Proposals to prevent the attacks include better network configurations and changes to protocol standards to close the integrity gaps.   

Our experimental section describes a MiM attack on a 4G network where all the over-the-air packets go through the MiM, but our MiM does not modify the contents of messages. We have also used the same open software and hardware described in the works \cite{adaptover_mim1,4G_5G} to show our detection mechanisms. While the authors of the above have focused on mechanisms to prevent the MiM attacks, our work can detect the MiM while the attack is happening. The goal of the REVEAL protocol is to detect MiM attacks irrespective of attack vector and wireless technology.

The MiM attack in SigUnder \cite{Sigunder} only modifies the un-encrypted broadcast messages of the 5G protocol. Since the MiM needs to first receive the broadcast messages before sending the modified contents, this half-duplex MiM can be caught by our detection methods. Taking advantage of weak authentication and lack of integrity in certain configurations of the 4G protocol, IMP4GT \cite{Rupprecht2020IMP4GTIA} MiM modifies the small sized uplink and downlink packets. The IMP4GT attack inserts a half-duplex relay between the base station and the mobile.  The REVEAL protocol can detect all of the above mentioned attacks as they use either half or full-duplex MiM devices.
\subsection*{Comparative Assessment: Reveal in Contrast to State-of-the-Art Approaches} 

Fingerprinting methods have become popular for MiM detection in the literature. Kim et al. \cite{RSS} have capitalized on variations in received signal strength before and after a MiM attack. However, the MiM can counter such detection mechanisms by adjusting its transmit power. Detection methods based on observing patterns in TCP-ACK pairs \cite{TCP_ACKs}, inter-packet arrival times \cite{IPAT}, round trip times \cite{RTT,RTT1}, and traffic \cite{traffic}, have also been proposed. However, radio propagation conditions can affect inter-packet arrival times, round trip times require accurate knowledge of the distance between the mobile and the base station, and user traffic affects TCP-ACK pairs. We intend for the REVEAL protocol to be a useful tool for detecting MiMs regardless of the attack vector or communication protocol employed.

Building on the work \cite{MiM_Kumar}, we present the REVEAL protocol and demonstrate how REVEAL exposes half, full and double full-duplex MiMs in a 4G network.  The method also works for WiFi and sub-6GHz 5G.

\section{Background}\label{background}
\subsection{Clock Synchronization}\label{clock-synch}

\begin{figure*}[t]
    \centering
    \includegraphics[width=.8\linewidth]{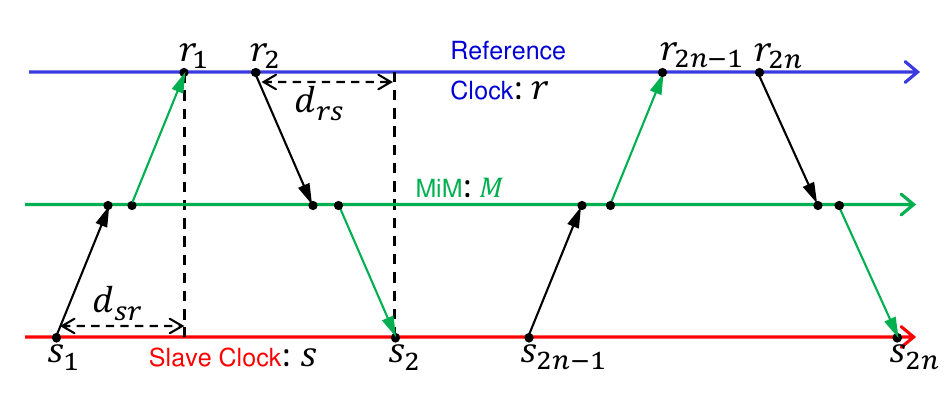}
    \caption{Clock synchronization in the presence of Man-in-the-Middle. (Both the durations $d_{rs}$ and $d_{sr}$ are in the units of the reference clock.) }
    \label{MiM}
\end{figure*}

Communicating devices share the wireless medium and must schedule message exchanges to operate in harmony.  For networked devices that share the wireless medium and use cellular technology for packet exchanges, the base station schedules both uplink and downlink traffic for the mobiles. To follow the schedules precisely the base station and mobiles must have the same time.

 
 Suppose the base station schedules a conversation from time $t_1$ to $t_2$.
 Then the mobile must use that opportunity to precisely initiate communication at $t_1$ and cease at $t_2$.  In 4G/5G network terminology, the base station grants the communicating opportunity, i.e., a time slot in a particular frame and using symbols $k_1$ to $k_2$.  If a mobile's timing is off, the window of operation can overlap with that of other devices, resulting in packet collisions which degrade network throughput.
 
 Indeed, the primary requirement put on the mobile to connect to scheduled communication technologies like cellular networks and to remain connected is to maintain time synchronization with the base station.
   As the time must be same for the entirety of the network, each mobile synchronizes its clock to the base station's clock.  Consider a base station that has a reference clock as shown in  Fig. \ref{fig:clock_relationship}, with the mobile's clock slave to that of the base station. The IEEE 802.11 family of technologies \cite{ieee_standards}, the cellular technologies 4G and 5G \cite{3gpp_specifications}, and the short range communication technology of Bluetooth \cite{bluetooth}, all  have distinct protocols to achieve clock synchronization. For ease of exposition, we present below only the fundamental idea behind synchronization. The documents \cite{4G_sync0,5G_sync} describe clock synchronization procedures in 4G and 5G.  
 
 Suppose $t_{ref}$ and $t_{slave}$ represent the clock-based time stamps from the reference and slave clocks respectively. If $t_{slave}=a*t_{ref}+b$ gives the relationship between the two clocks, then $a$ is the skew and $b$ is the offset of the slave. 
 The skew gives information on the rate of change of the slave clock with respect to the rate of change of the reference clock. Ideally, skew is 1. 
 The ambient temperature, humidity, pressure, vibrations, etc., affect the crystals in the clocks, and therefore the mobile needs to perform skew and offset estimation persistently \cite{sullivan1990characterization}.

To estimate skew and offset \footnote{The over-the-air clock synchronization mechanism is different in 4G/5G networks; however the core network follows a similar mechanism.}, the base station  and the mobile exchange packets that carry time stamps from their respective clock sources, i.e., the snapshot of the time just before sending a packet, and the snapshot of the time just after receiving a packet, both according to their local clocks. 
 Let $s_{m}$ denote the time stamp of a packet at the mobile, according to the mobile's clock, at the time it is sent, and let $r_{b}$ denote the time stamp at the base station when it is received, according to the base station's clock. For a packet travelling in the reverse direction, let $s_{b}$ denote the time stamp of a packet at the base station at the time it is sent, according to the base station's clock, and let $r_{m}$ denote the time stamp at the mobile when it is received at the mobile, according to the mobile's clock. Suppose $d_{mb}$ and $d_{bm}$ are uni-directional delays from the mobile to the base station and vice versa, measured according to the reference clock, which in this case we suppose to be that of the base station. Then $r_{m} = a_{m} s_{b} + b_{m} + a_{m} d_{bm}$
and $s_{m} = a_m r_{b} +b_m - a_{m}d_{mb}$.
To synchronize the clocks we need to estimate offset $b_m$, skew $a_m$, and also the path delays $d_{mb}$, $d_{bm}$. Suppose (for simplicity of notation) that even numbered packets $k,k+2, k+4, \ldots $ are sent from the base station to the mobile, while odd numbered packets $k+1,k+3, k+5, \ldots$ are sent in the reverse direction, as shown in Fig. \ref{fig:time_exchanges}. Then  
\begin{equation}
      \begin{pmatrix}
         r_{m,k} \\
    s_{m,k+1} \\
    r_{m,k+2}  \\
    s_{m,k+3} \\
      \end{pmatrix} = \begin{pmatrix}
              s_{b,k} & 1 & 0 & 1 \\
    r_{b,k+1} & 0 & -1 & 1 \\
    s_{m,k+2} & 1 & 0 & 1 \\
    r_{b,k+3} & 0 & -1 & 1 \\
      \end{pmatrix} \begin{pmatrix}
           a_m \\
    a_m*d_{bm} \\
    a_m*d_{mb}  \\
    b_m \\
      \end{pmatrix}
\end{equation}\label{eq1}
\noindent
This matrix is of rank 3 only, since the last column is the difference between the second and third columns. Hence, as pointed out in \cite{impossitility_theorem_proof}, in general it is impossible to estimate all four parameters $(a_m,b_m,d_{mb}, d_{bm})$. 
Thus clocks cannot be synchronized if the delays in the uplink and downlink directions are asymmetric. 
Consider, for example, a mobile for which skew = 1. If a packet transmitted at mobile time 100s is received at base station time 100.000001s, then it is impossible to distinguish whether delay $d_{mb}=0$ and $b_m=10^{-6}$, or $d_{mb}=10^{-6}$ and $b_m=0$.

\begin{figure*}[t!]
\centering
\includegraphics[width=.7\linewidth]{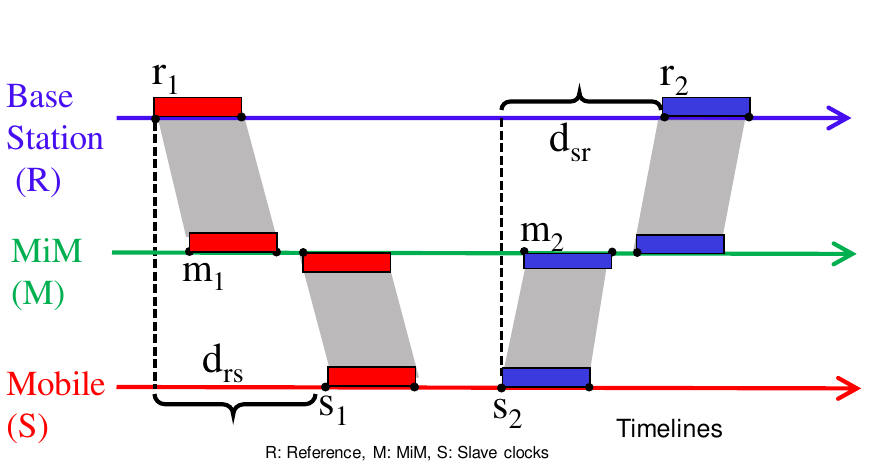}
\caption{Half-duplex middle node. (Both $d_{rs}$ and $d_{sr}$ are in the units of the reference clock, which in this case is the base station.)} \label{fig:half-duplex_block_diagram}
\end{figure*}

\begin{figure*}
\centering
\includegraphics[width=.75\linewidth]{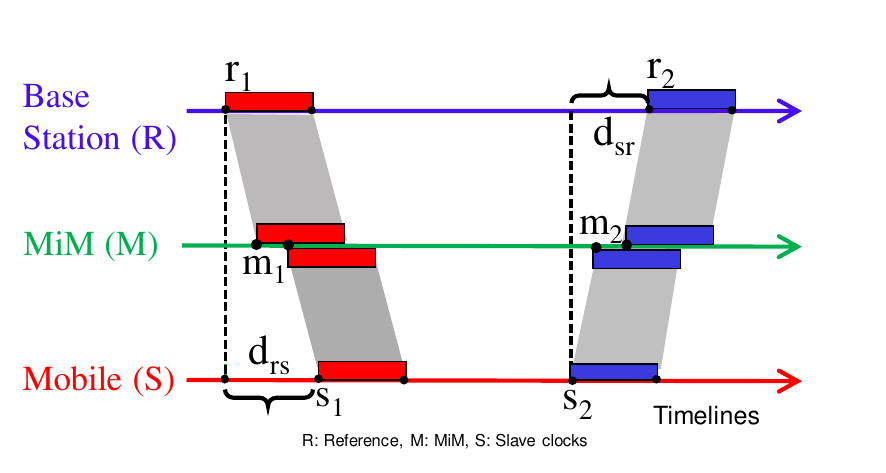}
\caption{Full-duplex middle node} \label{fig:full-duplex_block_diagram}
\end{figure*}
 
On the positive side, the
skew $a_m$ can be estimated by
\begin{equation}\label{skew}
    a_m=\frac{r_{2k+1}-r_{2n+1}}{s_{2k+1}-s_{2n+1}}.
\end{equation}
Another important capability is that for any packet transmitted by the mobile/base station, the mobile/base station can estimate the time at which it is received by the base station/mobile according to the base station's/mobile's clock \cite{impossitility_theorem_proof}. 
To see this, note that from (\ref{eq1}), 
$a_m*d_{bm}+b_m=r_{m,k}-a_m s_{b,k}$. Since all the quantities on the RHS are known or estimable, the quantity 
$a_m*d_{bm}+b_m$ can be determined based on time-stamped packet exchanges.
Hence for any future packet transmitted by the base station at local time $s_b$, its receipt time at the mobile, $r_m = a_m + s_b + a_m*d_{bm}+b_m$
can be determined by the base station.
That is, knowing $s_b$, the base station  can determine $r_m$. Likewise, knowing $s_m$, the mobile can determine $r_b$. 
This capability is used by REVEAL.

In the special case where the delays in the two directions are known to be equal, i.e., $d_{bm}=d_{mb}$, the delays $d_{mb} (=d_{bm})$ as well as offset $b_m$ can be estimated. The asymmetry in delays impact any Clock/Time synchronization protocols. The protocol Precision Time Protocol \cite{PTP} 
determines from technological considerations the amount 
$d_{bm}-d_{mb}$ by which the delays in the two directions are asymmetric. Subsequent to this, the remaining delays are symmetric in the two directions and can thus be estimated.  Extensive body of literature, challenges and solutions on time synchronization are compiled by the inventor of timing protocols, Dr. Mills \cite{Mills_ptp_ntp}.

When there is a MiM between the mobile and the base station, it can only delay the packets \cite{MiM_Kumar} from mobile to base station by a constant, say $d_{MiM,mb}$; else, if it delays different packets from mobile to base station by different times, then the inconsistent behavior can be detected. Likewise, the MiM can only delay packets from the base station to the mobile by a constant, say $d_{MiM,bm}$. 

A MiM interfering in any other way with the clock synchronization process itself, besides adding a constant delay in each direction,
can thereby be detected.
Based on this approach we synchronize  the clocks of the mobile and base station. We have conducted an experimental investigation and report on the estimation results 
in Section 
\ref{implementation}. 

\section{Capabilities of A MiM}
As noted above and shown in Fig. \ref{MiMs}, we can broadly classify an MiM node into three types based on its forwarding capability: Half-duplex , Full-duplex or Double Full-duplex. 

\textbf{Half-duplex MiM.} 
The half-duplex only capable MiM must first choose the direction of attack: uplink or downlink. 
Then, since simultaneous listening and forwarding is not possible for a half-duplex MiM, it must listen to the entire packet before forwarding. 
Hence, packets passing through a half-duplex node are delayed at least by the length of the packet. There may be an additional processing delay on top of this. This is shown in 
Fig. \ref{fig:half-duplex_block_diagram}.


\textbf{Full-duplex MiM.}
 As in the case of the half-duplex case, a full-duplex MiM must first choose the direction of attack: uplink or downlink. The concurrent reception and transmission of packets by a full-duplex MiM enables it to forward packets in the chosen direction
 quicker than its half-duplex counterpart, as illustrated in Fig. \ref{fig:full-duplex_block_diagram}.
 However, a limitation of a full-duplex MiM is its inability to manage simultaneous bidirectional traffic between the base station and the mobile. It can either forward packets from mobile to base station or base station to mobile, but not both at the same time.

\textbf{Double full-duplex MiM.}
A double full-duplex MiM is capable of listening to and forwarding traffic simultaneously in both directions, making it the ideal choice for orchestrating an attack vector and taking over the communication link between the base station and the mobile.

\begin{figure*}[t!]
\centering
\includegraphics[width=.7\linewidth]{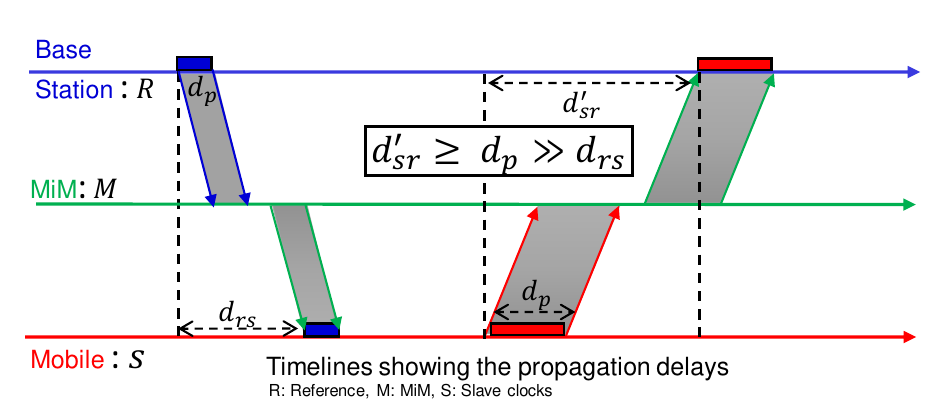}
\caption{Detecting half-duplex MiM} \label{fig:half-duplex-detect}
 \end{figure*}

 \begin{figure*}
     
 \centering
\includegraphics[width=.7\linewidth]{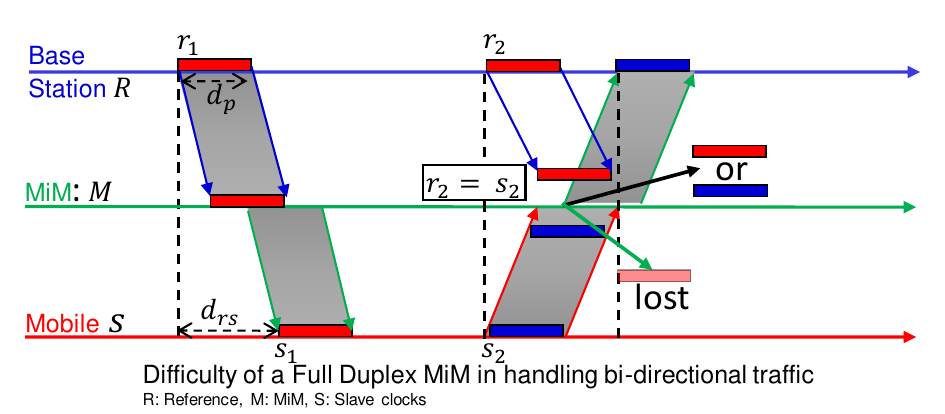}
\caption{Detecting full-duplex MiM} \label{fig:full-duplex-detect}

\end{figure*}
\section{Detection of MiM}\label{detection}
The REVEAL protocol is engineered to identify the existence of a Man-in-the-Middle (MiM) node, irrespective of its forwarding abilities. As outlined in Section IV, specific mechanisms customized for 4G/5G networks constitute fundamental components of the REVEAL protocol. Below, we detail the programming required for either the base station or the mobile device to detect the presence of a MiM. For a MiM with one of the three forwarding capabilities, we present detection state machines, and step by step implementation details.

\subsection{Half-duplex MiM}\label{half-duplex}
The half-duplex MiM must receive the complete packet before it can forward; therefore the delay is at least the duration of the packet itself. 
The transmission of messages through a half-duplex MiM therefore
results in significant delays, typically exceeding those caused by path delay, which is determined by the distance between the base station and the mobile. The additional delay is proportional to the length of the message, as shown in Fig. \ref{fig:half-duplex-detect}.
Hence long packets traversing through a half-duplex MiM experience much higher delays.  
\begin{figure}[H]
    \centering
    \includegraphics[width=.8\linewidth]{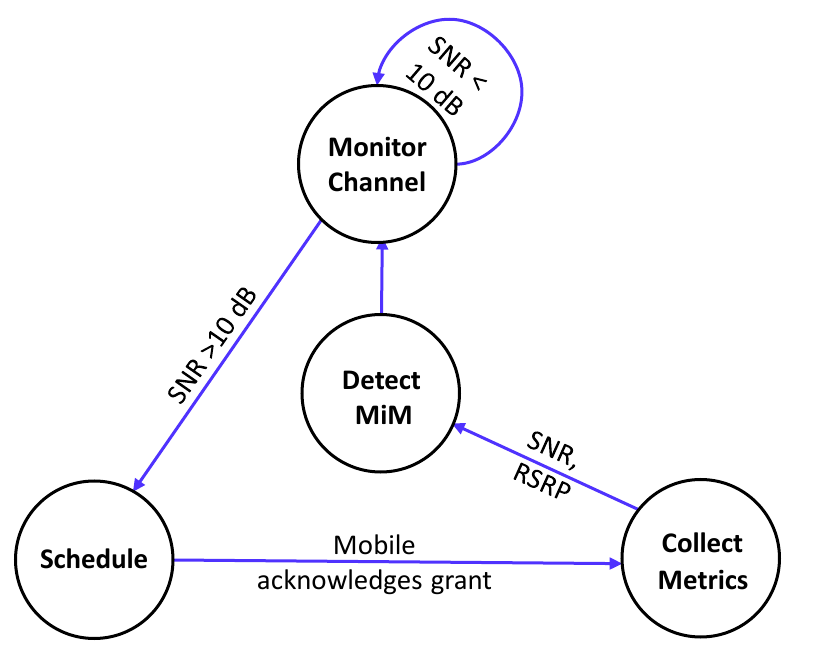}
    \caption{State machine: Half-duplex MiM detection}
    \label{fig:statemachine_hdd}
\end{figure}

Detecting a half-duplex MiM can be achieved by sending a sufficiently long packet. 
Fig. \ref{fig:statemachine_hdd} presents the state machine of REVEAL protocol to expose a half-duplex MiM.
The protocol instructs the base station to schedule a long packet for the mobile. To avoid fragmentation errors, protocol may send a burst of shorter packets back to back. 
The MiM is forced to add an additional delay
at least equal to the increase in the length of the packet.
 The REVEAL protocol capitalizes on this fact to detect the presence of a half-duplex MiM, as described below. 
 
 The base station monitors the channel conditions in its ``Monitor channel" state in the state machine of the protocol in Fig. \ref{fig:statemachine_hdd}.

 Cellular standards \cite{5gcqi} provide the required signal thresholds for data decoding. The Channel Quality Indicator \cite{5gcqi} provides feedback on channel conditions that inform either the base station or the mobile to choose the ideal modulation and coding level for data transfer. 
 In a 4G/5G network, when the link signal-to-noise ratio (SNR) is 10 dB and above, up to 64 QAM modulated symbols pass through the wireless link with high probability \cite{ghosh2011essentials}\footnote{Also, experimentally we have found that detection tests have very good accuracy at SNR above 10 dB, further reference in Section \ref{implementation}}. At such high SNR levels there is almost no packet loss and therefore there is no additional delay in reception of the sent packets. Our detection method challenges the half-duplex MiM by scheduling long packets, with the half-duplex MiM having to listen to the entire packet before forwarding, which induces significant packet delay.
 
 The base station continues to monitor the SNR and moves to the ``Schedule" state when SNR $\geq$ 10 dB. In this state, the base station gives a long communication opportunity to the mobile either in downlink or uplink. The minimum scheduling opportunity in a 4G cellular network, transmission time interval (TTI), is 1 millisecond. To schedule longer opportunities, the base station allocates continuous communication opportunities. After receiving confirmation from the mobile that it is aware of the granted opportunity, the base station moves to the ``Collect metrics" state. In this state, the base station records the SNR and received power. The half-duplex MiM is unaware when the data transmission will end, and continuously listens before forwarding. 

Meanwhile, while waiting for the data from the MiM in the scheduled opportunity, the receiver experiences a silent period as the messages are delayed beyond the scheduled time.  In this silent period, received power and SNR are poor. 
 
 The base station collects the metrics, SNR and received power, and moves to the ``Detect MiM" state, where it decides whether a MiM is present. 

\subsection{Full-duplex MiM}
A full-duplex MiM can forward incoming packets as soon as they arrive, but can do so only in one direction at a time. Consider a downlink packet that is processed by the MiM and transmitted towards the mobile, as shown in Fig. \ref{fig:full-duplex-detect}. 
The base station and mobile coordinate to transmit two packets, one transmitted by the base station at time $r_2$, and one transmitted by the mobile at time $s_2$. These packets are timed so that they impinge on the MiM at the same time. Hence the MiM cannot simultaneously forward both, and is forced to abandon packets in one of the directions. By checking if they have received each other's packets, the base station and mobile can detect the failure of the MiM to forward both packets.  

Achieving this coordination depends critically on the results of the Clock Synchronization algorithm. As shown in Section \ref{clock-synch}, a mobile interfering in any way other than adding a constant delay to packets in a direction, can be detected.  
As noted in Section \ref{clock-synch}, an outcome of the clock synchronization process is that the mobile can time its packet transmission so that the MiM-relayed packet is received by the base station at a specified time according to the base station's clock. Similarly, the base station can time its packet transmission so that the MiM-relayed packet is received at a specified time by the mobile according to the mobile's clock. Using this capability as well as the choice of the length of the packets the base station and mobile can ensure that a conflict is created at the MiM. Therefore, this precisely timed bi-directional data traffic test exposes the presence of a Full-duplex MiM. 

\begin{figure}[h]
    \centering
    \includegraphics[width=.8\linewidth]{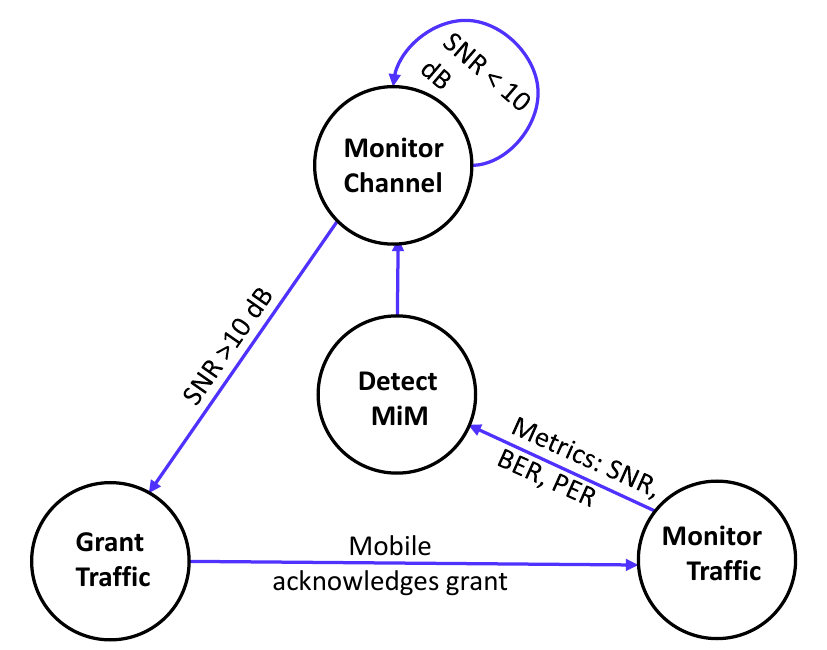}
    \caption{State machine: Full-duplex MiM detection}
    \label{fig:statemachine_fdd}
\end{figure}

\begin{figure}
    \centering
    \includegraphics[width=.8\linewidth]{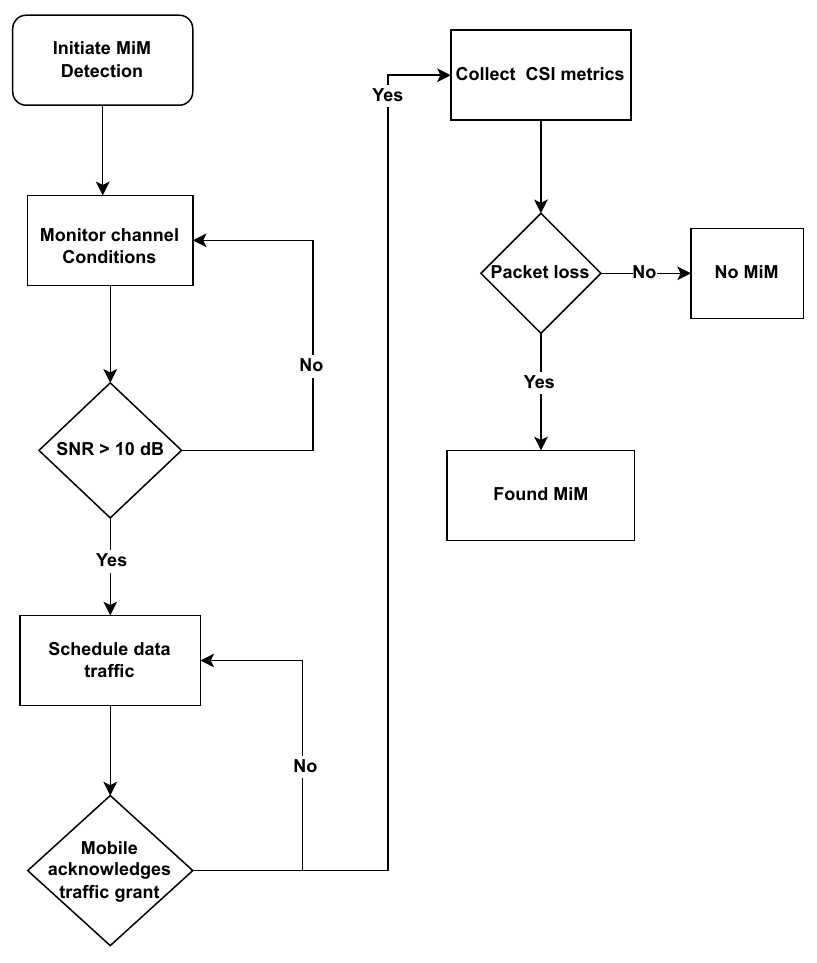}
    \caption{Protocol flowchart}
    \label{fig:flowchart}
\end{figure}
The state machine for the REVEAL protocol which is used to detect a full-duplex MiM is depicted in Figure \ref{fig:statemachine_fdd}. As described in Section \ref{half-duplex}, there is an initial period of monitoring.
After this, the base station proceeds to the ``Grant traffic" state, where it schedules uplink and downlink traffic simultaneously. Once the mobile acknowledges the receipt of these grants, the base station moves to the ``Monitor traffic" state to monitor the metrics of SNR, bit error rate (BER), and packet error rate (PER). These metrics indicate whether the packets reach their intended destination. After collecting the metrics, the base station enters the ``Detect MiM" state, where it determines whether a MiM is present. If a MiM is detected, the base station moves back to the ``Monitor channel" state to observe the channel conditions again.
The flow chart describing the detection protocol is shown in Fig. \ref{fig:flowchart}

\subsection{Double full-duplex MiM}
A double full-duplex MiM can forward traffic in both directions simultaneously, and can do so with little delay. The time-driven conflicts that the REVEAL protocol uses to expose half and full-duplex MiMs cannot detect a double full-duplex MiM. However, we show how wireless technologies that are capable of operating on multiple frequency bands can still detect the presence of a double full-duplex MiM.
Cellular technologies like 4G and 5G operate in frequency-division duplex mode \footnote{Our work can  be extended to the time-division mode of cellular networks, but for clarity of presentation we focus only on the frequency-division duplex mode.}, and both the base station and the mobile are designed to operate on several bands for interoperability and seamless operation across the globe. 

\begin{figure}
        \centering
    \includegraphics[width=.7\linewidth]{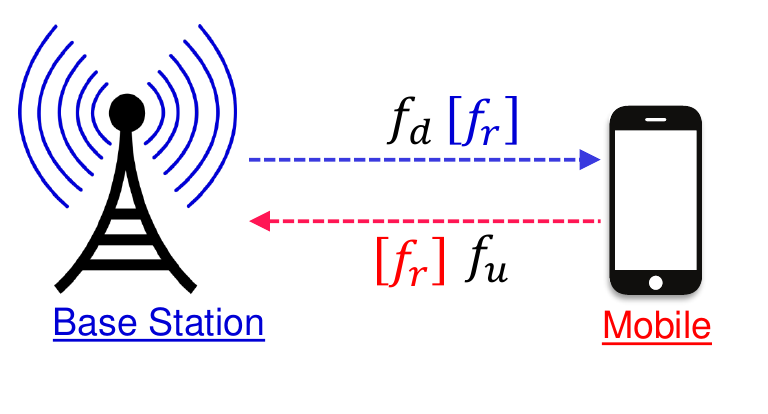}
    \caption{Detecting double full-duplex MiM}
    \label{two_full_duplex_detect}
    
\end{figure}

As depicted in Fig \ref{two_full_duplex_detect}, a frequency-division duplex (FDD) base station uses frequency $f_d$ to communicate with the mobile, and the mobile uses a frequency $f_u$ to talk to the base station. A double full-duplex MiM listens and forwards transmissions on both $f_d$ and $f_u$. To attack the FDD link, the MiM needs knowledge of the operating band. The MiM can scan the spectrum to detect transmissions in $f_d$ and $f_u$. After narrowing down on the frequencies, the double full-duplex MiM launches the attack. 

The REVEAL protocol takes advantage of the need for the double full-duplex MiM to perform spectrum sensing 
in order to launch the attack. The base station can instruct a mobile via an encrypted control message asking it to change its frequency of operation in either downlink or uplink frequency to $f_r$. Unaware of this change, the double full-duplex MiM still listens and forwards data on prior sensed frequencies. Consequently all the packets that are communicated on $f_r$ link are lost. 
One choice of $f_r$ could be such that $ |f_d-f_r| > BW$ or $ |f_u-f_r| > BW$, where BW is the sensing bandwidth of the MiM. The ideal choice of BW is in the order of GHz. \footnote{Commercial 4G/5G cellular modems have operating bandwidth anywhere from 20 to 400 MHz.} A large BW forces the double full-duplex MiM to have additional capability of wider operating bandwidth $(BW_{op})$. Spectrum sensing takes a longer duration if  $BW_{op} << BW$.  

\begin{figure}
    \centering
    \includegraphics[width=.8\linewidth]{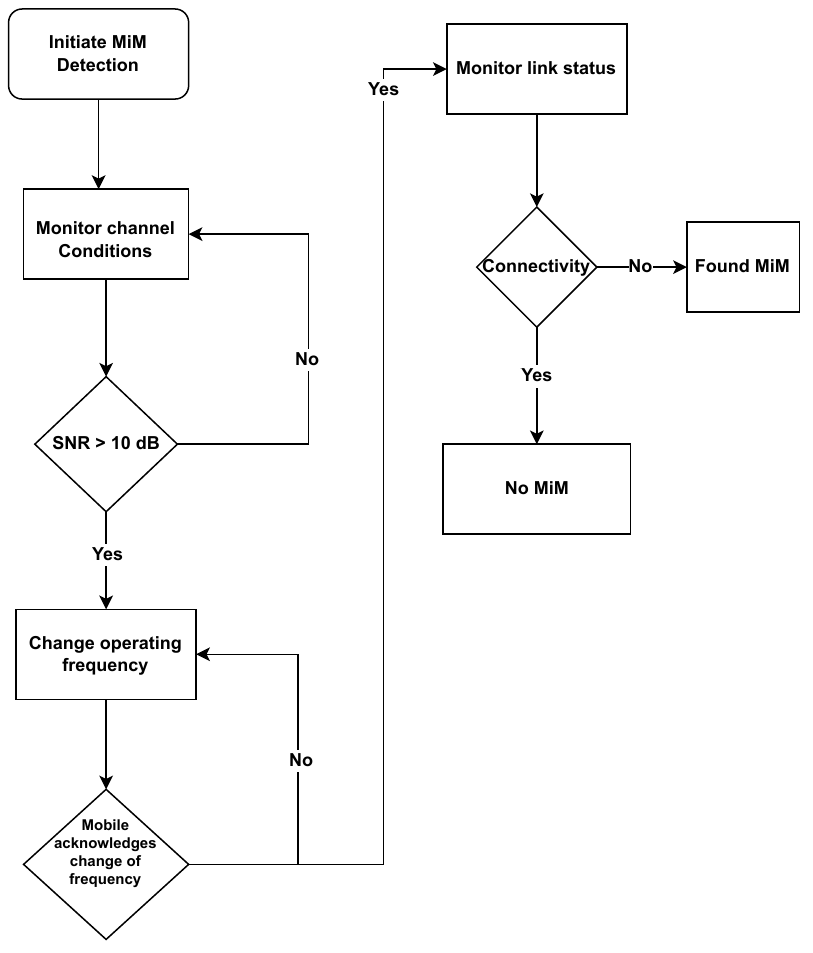}
    \caption{Flowchart for double full-duplex MiM detection}
    \label{fig:flowchart_double}
\end{figure}
To detect a double full-duplex MiM, REVEAL follows the sequence of steps in Fig. \ref{fig:flowchart_double}. As and when the base station decides to perform a frequency change test, and channel conditions (SNR)  are good, the base station informs the mobile to change either uplink or downlink frequencies, or both. Once the base station receives acknowledgement from the mobile indicating the receipt of the message, the protocol monitors the link state and observes the radio conditions on the new frequencies. 

The base station detects the MiM by a failure to receive the packets sent by the mobile on the new frequencies, as the transmissions must go via the double full-duplex MiM to  reach the base station. Since our detection mechanism relies packet reception failure, we need to perform test when SNR is good, to ensure packet reception failed only because of presence of an MiM. Experimently, we found that at SNR $\geq$ 10 dB, there are almost no packet reception failures, which is an ideal condition to perform our detection test. The SNR and received power estimates during the detection test indicate whether there is an ongoing transmission. Observing the mobile's connection status,  the base station decides whether a double full-duplex MiM is present or not.

\subsection{REVEAL Protocol}
    The goal of the REVEAL protocol is to detect a man-in-the-middle, whether it is half, full, or double full-duplex capable. Therefore, the protocol should independently move to MiM detection mechanisms. In the detect state of Figs. \ref{fig:statemachine_hdd},\ref{fig:statemachine_fdd}, and \ref{two_full_duplex_detect}, we perform detection steps mentioned in Section IV. The REVEAL protocol schedules detection methods with the scheduling policy left open. The system designer can follow the scheduling policy of his/her choice, following which the REVEAL protocol moves to any of the state machines presented in Figs. \ref{fig:statemachine_hdd}, \ref{fig:statemachine_fdd}, and \ref{fig:flowchart_double}. There is no particular policy that has an advantage over another as a MiM attack can happen at any point in time. 

\section{Implementation}\label{implementation}

\subsection{4G Network}
We use srsRAN 4G network source code \cite{srsran}, a public software project that has 4G base station (eNodeB), mobile (user equipment, UE) and network core. Using srsRAN, a host machine with USRP X310 \cite{usrpx310} as base station and another host machine with USRP B210 \cite{usrpb210}, we create a local 4G network and show attack and defence mechanisms.

On machine 1, we launch network core and eNodeB softwares. Machine 1  connects to a USRP X310. eNodeB software interfaces with USRP X310 and acts as base station. Similarly, on machine 2, we run UE software that interfaces with USRP B210. This combination makes a 4G mobile. The mobile, once connected to the 4G network successfully, gets an IP address assigned. Any application on machine 2 can then communicate with machine 1 via 4G network. 

\begin{figure*}[t]
\hspace{1.5cm}
    \centering
    \includegraphics[width=1\linewidth]{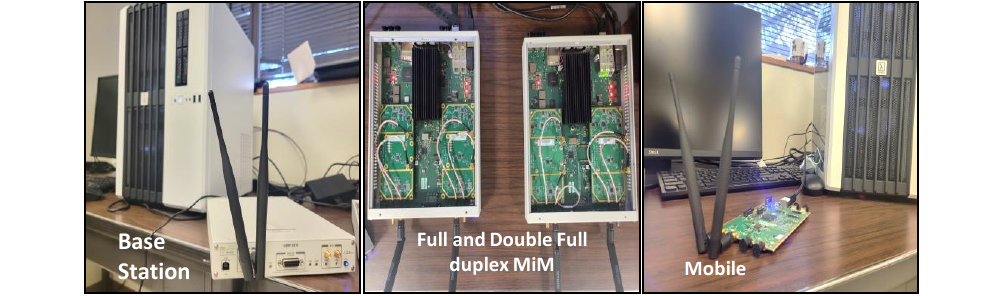}
    \caption{Software defined radio 4G Network and MiM used for experiments}
    \label{fig:setup}
\end{figure*}

 We operate both uplink center carrier at 2400 MHz and downlink center carrier at 2500 MHz, on ISM frequencies. Nevertheless, upon procuring appropriate licences to use Cellular operator's frequencies, our experiments can be reproduced on a live network too.  When connected directly to the base station in the absence of MiM, at the mobile side, we observe  pathloss of 64 dB, reference signal received power of -64 dBm, timing advance (a metric that the 4G protocol uses to adjust the timing offset between the base station and the mobile) to be 0.5 $\mu$S, SNR as high as 20 dB, and Modulation and Coding Scheme index (a parameter that indicates channel conditions for data scheduling \footnote{The higher the better; the maximum usable index is 28.}) of 28, during a  bi-directional traffic session.  We configure RF gains of eNodeB and UE conservatively to ensure there are no packet losses either in uplink or downlink. The locations of the mobile, the base station, and the MiM remain the same throughout in a particular trial of our experiment. Using the clock synchronization method  in Section \ref{background} and assuming bi-directional delays are same, employing Eqn. \ref{skew}, we found that the skew between the base station and the mobile clock is 0.9935607. Also, since REVEAL protocol doesn't have any dependency on Cellular technology, using open-source 5G software like Open Air Interface 5G project \cite{OAI-5G}, and srsRAN 5G project \cite{srsran5G}, our results can be reproduced on 5G network as well. Our software might be helpful in debugging and to quickly reproduce results \cite{MIM_UE,Time_sync_scripts1}. A short demo version of our work is available \cite{mim_demo}. The distance UE and eNodeB is varied from 2.5 to 8 mts in our experiment trials. We have repeated all the experiments on the licensed frequencies to show that our work can immediately transferable to the current 4G/5G networks. Since, we have performed controlled experiments on licensed frequencies, we have highlighted them outside of the current in the github link \cite{demos}.

\subsection{Full-Duplex MiM}
Constructing wireless transceivers capable of both transmitting and receiving on the same channel, known as full-duplex transceivers, represents a significant technical challenge due to the complexity of the required radio hardware architectures \cite{full_duplex_arc}. The successful operation of such a full-duplex transceiver is contingent upon the use of both digital and analog signal cancellation to achieve over 100 dB isolation between the transmitter and receiver.

Upon conducting a thorough investigation of the cellular standards \cite{3gpp_specifications}, we have ascertained that there is no explicit verification mechanism for carrier frequencies. Operators of cellular networks are mandated to comply with the guidelines stipulated by local regulatory authorities pertaining to the utilization of radio frequencies. Consequently, cellular protocols have been designed to allow the base station and mobile to communicate across various frequency bands. Additionally, the initial step for mobile communication is the detection of a cellular base station irrespective of the network operator, following which a connection request is initiated with a ``visible base station". Subsequently, access to the network is granted after successful validation of access credentials.
 
Such a design choice in cellular standards greatly simplifies the hardware architecture of a full-duplex MiM. By allowing the base station and the mobile to communicate on a multitude of frequency bands, the full-duplex middle node can listen to the base station's downlink traffic on frequency $f_d$ and forward it on a different frequency $f_m$. As a result, the mobile can discover the base station on $f_m$ instead of $f_d$.

\subsubsection{Full-duplex MiM prototype}
We use an additional USRP X310 to build the equivalent of a full-duplex MiM. The X310 radio has two independent radio chains and can simultaneously receive and transmit. We configure one of the radio chains to receive the base station transmissions and another chain to re-transmit the received samples. We do not modify the received samples. In our experiment, as we are aware of our 4G/5G base station's operating frequencies, we directly configured the MiM to receive on those frequencies.  With any open-source radio software like GNU-Radio, it is also straightforward to develop a spectrum analyzer, or using public databases \cite{cell_towerdatabase} it is fairly easy to identify the operating frequencies of a target base station. 

  A full-duplex MiM need not listen to the entire transmission before forwarding. It can choose the forwarding delay to be as small as possible. The receive chain of X310 has a digital down-converter to collect the base-band samples. These samples are then sent to the digital up-converter of the transmit chain of X310. We configure the X310's receive chain to stream packets of base-band samples to the transmit chain. The time to down-convert the over-the-air signal, length of the streaming packet and the time to up-convert to a different frequency comprise the total additional delay our MiM adds to the link.  To reduce transit time of samples through the MiM, the only controllable parameter in X310 is the size of the streaming packet. We configure the samples per packet to its least possible value; each packet carries 24 of 32 bit base-band samples from receive chain to transmit chain. The major one would encounter is to approriately adjust the transmit and receive power of the X310. We found a good configuration of transmit gain and receive gain for our full-duplex MiM so that it would saturate the signal either at eNodeB or UE. In our experiments where the distance between eNodeB and UE is upto 8m, we found 30 dB of TX and RX gain is sufficient. The source files of the MiM and instructions on how to produce binaries and operationals aspects of MiM are available at \cite{MiM_Full_Duplex_Implementation}.

\subsubsection{Full-duplex MiM attack}\label{4g_MiM}
Since we have access to the configuration files of the base station and the mobile, we can operate the 4G network on frequencies in the ISM band and with operating bandwidth of 10 MHz. Also, our configuration ensures there is no direct link between the base station and the mobile. The 4G base station and the mobile operate in frequency-division duplex mode, i.e., they use different radio frequencies for uplink $f_u$= 2400 MHz and downlink $f_d$= 2500 MHz. Our goal is to create a full-duplex MiM and show that the mobile connects to the 4G network via MiM, but not directly to eNodeB. Our 4G  MiM listens to $f_d $= 2400 MHz and transmits the incoming packets without changing the contents on $f_m $= 2500 MHz. On machine 2, we configure the mobile to listen downlink on $f_m$ = 2500 MHz, and transmit uplink on $f_d$ = 2400 MHz. Because of this configuration, the UE cannot directly listen to eNodeB. All the communication to the UE must go via the MiM and not directly from eNodeB as the base station's, and mobile's operating frequencies are configured to listen to the MiM. In this particular configuration MiM's preference is forward any traffic on downlink of eNodeB. The full-duplex MiM may also choose to attack uplink. 

As shown in Fig. \ref{fig:setup}, we keep eNodeB and UE 2m apart and keep MiM in between them.
Our first observation is that the mobile naively connects to the network and gets an IP address from the network. This proves that our full-duplex MiM is operational. The SNR at UE is 18 dB, and pathloss is 64 dB. Since the communication is via MiM, it is delayed. eNodeB indicates to the UE that the timing adjustment value i.e., the calculation made from the round-trip-time as part of initial access procedure is 5.2$\mu S$, whereas it is 0.5 $\mu S$ when UE is directly connected. As the UE can be located anywhere in the coverage area of eNodeB, only an Oracle can locate the UE; therefore the timing adjustment value does not help in detecting the MiM. The clock skew between the eNodeB and UE in the presence of full-duplex MiM is 1.002443. 

\subsubsection{Detection}
\begin{figure}
    \centering
    \includegraphics[width=.8\linewidth]{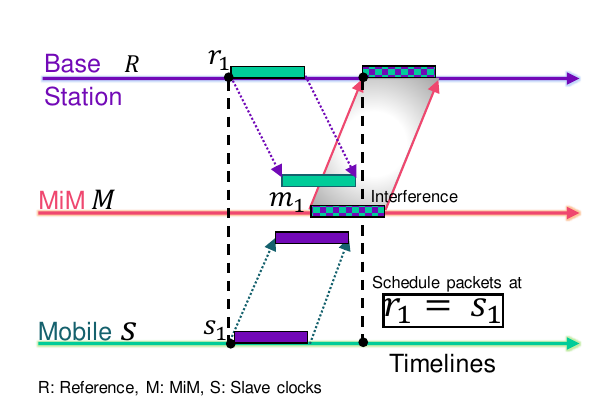}
    \caption{Block diagram indicating packet interference that happens in the presence of Full-duplex MiM}
    \label{fig:Packet_timeline}
\end{figure}

To detect the MiM, it is sufficient to initiate bi-directional traffic between the base station and the mobile.  Fig. \ref{fig:Packet_timeline} shows the mechanics of the detection. The 4G MiM is forced to make a choice, either to forward uplink traffic or to forward downlink. The packet in one of the two directions that is dropped at the 4G MiM cannot reach its destination, exposing the presence of the MiM. Since a 4G network operates in frequency-division duplex mode, the 4G MiM forwards traffic on uplink frequency or downlink frequency.

Frequency configuration in our network is slightly different from the standard configuration as we configure the uplink transmit frequency of UE to be the same as the downlink transmit frequency of eNodeB, which usually are different. This network configuration allows all the uni-directional traffic to reach its destination. However, when the eNodeB and UE transmit at the same time, the packets naturally collide at the MiM. Instead of looking for a lost packet during the MiM detection test, we can observe the SNR of the link to detect the presence of the MiM since the SNR at both the mobile and the base station drops due to the collision, ideally to 0 dB.

\begin{figure*}[h]
\begin{subfigure}[b]{0.5\textwidth}\centering

    \centering
    \includegraphics[width=.7\linewidth]{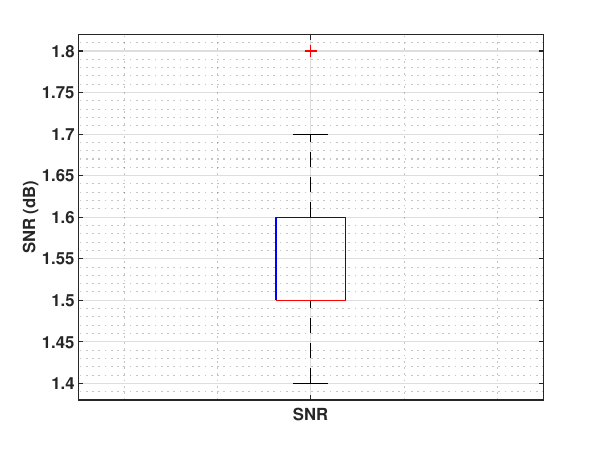}
    \caption{ SNR}
    \label{fig:snr}
\end{subfigure}
\begin{subfigure}[b]{0.5\textwidth}\centering
    
    \centering
    \includegraphics[width=.7\linewidth]{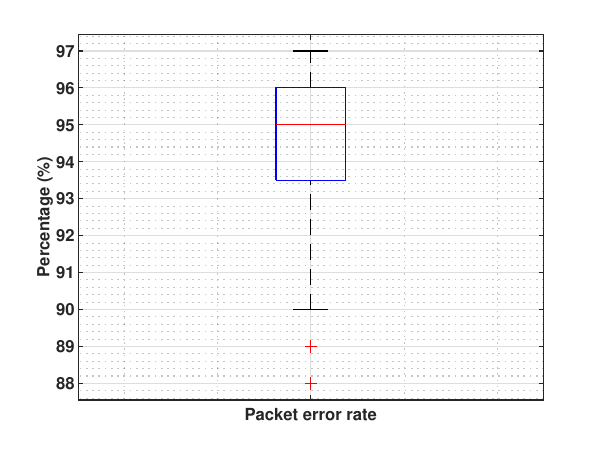}
    \caption{Packet error rate}
    \label{fig:per}
    \end{subfigure}
\caption{Experiments of several trials indicate the high performance of detection protocol}
\end{figure*}

In one of the detection trials, the SNR in the downlink was 14 dB at the mobile side from transmission time units 996 to 971. During this time, packets were only sent in the downlink from the eNodeB to the UE. From transmission time unit 972 until 975, bi-directional traffic was initiated, with packets being sent in both directions. Bi-directional packets from 971 to 975 arrived at the 4G MiM at the same time and on the same frequency, and the MiM was unable to separate these packets. As a result, the eNodeB and UE received collided packets. Since two packets of the same energy collide, the SNR of the packet is ideally 0. However, in our case, we observed an approximate SNR of 1 dB. We repeated this detection experiment 50 times, and the box plot of SNR at the time of detection is shown in Figure \ref{fig:snr}, with a median SNR of 1.5 dB. Additionally, Figure \ref{fig:per} shows that the median packet error rate over the trials is 95\%. 

In Table \ref{tab:full-duplex}, we tabulated the key measurements that show our tests are able to detect MiM with highest accuracy. The Uplink (UL) data channel SNR measured at eNodeB shows that there is interference on the channel. The control channel SNR shows in trial of experiments, we expect it to be high and reflection actual channel condition. Power head room (PHR) is a 4G metric that indicates whether UE is transmitting with power closer to max allowed power limits. Higher values indications, transmit power is far from maximum. However, in case of consistent UL transmission failures, UE transmits at higher power, in which case power head room is zero. Buffer status is the amount of data waiting at UE to be delivered to eNodeB. The Table\ref{tab:full-duplex} shows 10 trials of detection tests each lasting a second. In each trail, broadly, following are our observations. The UL data channel SNR is closer to 1 dB, good control channel SNR, in most of the trials, UE is transmitting with maximum power indicating significant packet failures, and close to 100 $\%$ packet error rate. The experiments are repeated contiguously, therefore, the packet buffer of UE remained the same, which indicates consistent failures throughout the detection tests. Negative acknowledgements and bit rate show that no or tiny amount of traffic is delivered to eNodedB from UE. Overall, we observe the consistency of detection tests as packets from UE fail to reach destination even with the least MCS. When MCS is 1, UE is using QPSK modulation, and code rate approximately .1, which implies that even under edge radio conditions our method has the highest accuracy of detection a full-duplex MiM. 

Suppose, the full-duplex MiM prefers to forward only uplink traffic when it receives both uplink and downlink transmissions at the same time. Cellular systems 4G and 5G are scheduled and base station driven technologies. The eNodeB schedules both the downlink and uplink traffic resources for a UE. The moment a full-duplex MiM starts abandoning downlink transmissions and forwards only uplink data traffic, neither of eNodeB's downlink traffic reaches UE nor the scheduling grants for uplink traffic, which are also on the downlink reaches UE. Such scenario is shown in Table  \ref{tab:full-duplex1}. Channel quality index (CQI), a 4G metric refers to channel conditions as measured by the UE and ranges from 0-15, higher the better. In our experiment trial, CQI is 9, indicating downlink SNR is $>$ 10 dB. Table \ref{tab:full-duplex1} shows the channel conditions collected at the eNodeB while detection tests are ongoing in 6 different sessions. In each of the trials, schedules for the uplink traffic are not available to UE as the MiM is forwarding only uplink traffic every time it encounters simultaneous traffic from both eNodeB and UE. Therefore, eNodeB has not received acknowledgements from the UE for most of the packets sent, hence PER is around 92$\%$. Also, UE channel measurements are not able at the eNodeB, since UE did not receive any uplink traffic grants. Table \ref{tab:full-duplex1} represents anomalous channel conditions which happen only in the presence of a full-duplex MiM.

\begin{table*}[]
    \centering
    \begin{tabular}{||c|c|c|c|c|c|c|c|c||}
    \hline
UL Data  SNR & UL Control SNR & PHR Value &MCS&Bit Rate (bps) &Acks &No Acks&PER ($\%$)&Buffer status \\
\hline
\hline
1.4&13.0&40&1&1.7k&1&59&98$\%$&8.61k\\
1.0&7.3&0&1&0&0&860&100$\%$&8.56k\\
1.0&7.6&0&1&0&0&856&100$\%$&8.61k\\
1.0&8.5&0&1&0&0&851&100$\%$&8.56k\\
1.0&9.2&0&1&0&0&852&100$\%$&8.61k\\
1.0&11.3&40&1&8.7k&5&746&99$\%$&8.61k\\
1.0&10.0&0&1&0&0&861&100$\%$&8.67k\\
1.3&9.8&40&1&1.3k&1&607&99$\%$&0.0\\
1.2&10.1&40&1&1.7k&1&790&99$\%$&8.56k\\
     \hline
    \end{tabular}
    \caption{Full-duplex MiM detection: Link layer measurements, MiM prioritizes downlink traffic}
    \label{tab:full-duplex}
\end{table*}

\begin{table*}[]
    \centering

\begin{tabular}{|c|c|c|c|c|c|c|c|c|c|}
\hline
\multicolumn{6}{|c|}{eNodeB-Downlink} &\multicolumn{4}{c|}{UE-Uplink}\\
\hline
\hline

CQI &MCS&  Bitrate&ACK & No ACK& PER($\%$) & Data SNR & Control SNR & Bitrate & Buffer\\
\hline
 9&1& 21k&12 & 151 & 92$\%$ &\multicolumn{2}{c|}{N/A}  & 0& 0.0\\
 9&1& 22k&13 & 156 & 92$\%$ &\multicolumn{2}{c|}{N/A}   & 0 &0.0\\
10&1& 20k&12 & 160 & 93$\%$ &\multicolumn{2}{c|}{N/A}   & 0& 0.0\\
 9&1& 21k&13  &155 & 92$\%$ &\multicolumn{2}{c|}{N/A}   & 0& 0.0\\
 9&1& 21k&14 & 156 & 91$\%$ &\multicolumn{2}{c|}{N/A}   & 0& 0.0\\
 9&1& 21k&14  &153 & 91$\%$ &\multicolumn{2}{c|}{N/a}    & 0& 0.0\\
 \hline
\end{tabular}
  \caption{Full-duplex MiM detection: Link layer measurements, MiM prioritizes uplink traffic}
    \label{tab:full-duplex1}
\end{table*}
\begin{figure}[h]
    \centering
    \includegraphics[width=.8\linewidth]{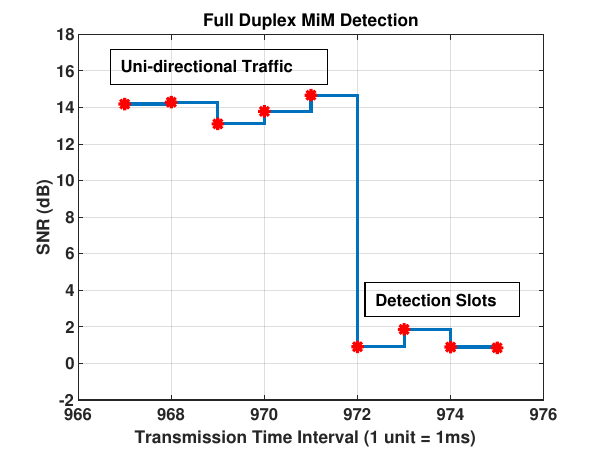}
    \caption{4G MiM detection experiment: Plot shows SNR at the mobile in each transmission time interval. From slots 966 to 972 mobile receives uni-directional traffic, but later when bi-direction traffic is initiated, SNR is poor due to interference, indicating presence of a MiM}
    \label{fig:4g_MiM_detect_experiment}
\end{figure}

\subsection{Double Full-Duplex MiM}
A double full-duplex MiM forwards the traffic in both directions. We reuse the design of the 4G MiM described in Section \ref{4g_MiM}. A full-duplex MiM only forwards traffic in one direction, but a double full-duplex MiM forwards traffic in both directions. So, we use two USRP X310s in conjunction to build a double full-duplex MiM. 

We operate the 10 MHz bandwidth 4G base station  on frequencies in ISM band, downlink $f_d$= 2400 MHz, and uplink $f_u$= 2500 MHz. The mobile listens on $f^{'}_d$= 2440 MHz and transmits on $f^{'}_u$= 2480 MHz. The double full-duplex MiM listens on $f_d$= 2400 MHz and forwards downlink traffic on $f^{'}_d$=2440 MHz. To forward uplink traffic, double full-duplex MiM listens on $f^{'}_u$=2480 MHz and forwards on $f_u$ = 2500 MHz.

Again, the network configuration ensures that the mobile can listen to the base station's traffic only via the MiM. Similar to the full-duplex MiM experiment, the mobile connects to the  network and receives an IP address. The clock skew between the  base station and mobile in this experiment is found to be 1.003142. 

\begin{figure}[h]
    \centering
    \includegraphics[width=.8\linewidth]{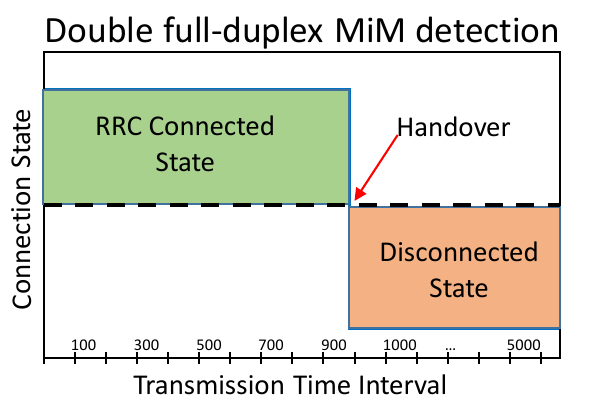}
    \caption{The mobile is connected to the base station untill TTI 900, after which mobile disconnects as there is no communication available on the carrier frequency}
    \label{fig:two_full_detect_4g}
\end{figure}

\subsubsection{Detection}
To detect a double full-duplex MiM node, the REVEAL protocol requires the base station to change its frequency of operation. In a 4G network, changing the operating frequency can be achieved by requesting the UE to perform a handover to a new eNodeB. Handover involves changing the mobile device's connection to a neighboring base station that operates on a different carrier frequency. When the mobile device receives the hard handover trigger, it searches for the new base station. We assume that the double full-duplex MiM cannot decode the encrypted messages sent from the base station that request the mobile device to perform a handover. As the MiM is unaware of the handover trigger, it continues to listen to the previous base station. 

To perform the attack, the MiM also needs to ensure that there is no direct link between the base station and the mobile device. To achieve this goal, the MiM can either operate on different frequencies or transmit at a much higher power to attack a mobile device that is far away from the base station. Since the MiM is unaware of the change in frequency and since the neighbor base station cannot reach the mobile device directly, the mobile device loses connectivity after attempting to handover. The ``Radio Resource Configuration (RRC)" state is used to monitor the ``Radio connected" state of the mobile device. When the mobile device has ongoing data traffic, the ``RRC" state remains in the ``Connected" state. The normal ``RRC" state toggles between Idle and Connected. As shown in Figure \ref{fig:two_full_detect_4g}, prior to handover, the mobile device is in the ``Connected" state. When attempting to handover to a ghost base station, the mobile device disconnects from the base station and remains in that state for extended periods.
The repository \cite{demos} maintains video demonstrations of the MiM attacks and detection process of REVEAL protocol.

\section{Discussion and Conclusion}
\label{sec:conclusion}
In this study, we have proposed the REVEAL protocol as an agnostic wireless man-in-the-middle detection mechanism. Through our experiments, we have demonstrated that REVEAL is capable of detecting half, full, and double full-duplex man-in-the-middle attacks. The detection protocol can be run continuously to monitor the network for potential attacks. Since, a potential attack on network can happen at anytime, strictly speaking there may not be an optimal detection scheduling policy other than continuous network monitoring. However an operator may choose a policy based on criticality of network under management. Network operators can integrate the REVEAL protocol into their traffic scheduling policies to enhance network security and prevent unauthorized access.

\textbf{Acknowledgements}
This material is based upon work partially supported by the
US Army under W911NF-22-1-0151 and W911NF2120064,
US Office of Naval Research under N00014-21-1-2385, and
Department of Homeland Security under 70RSAT20CB0000017.

\bibliographystyle{ieeetran}
\bibliography{bib}

\begin{thebibliography}{10}
\providecommand{\url}[1]{#1}
\csname url@samestyle\endcsname
\providecommand{\newblock}{\relax}
\providecommand{\bibinfo}[2]{#2}
\providecommand{\BIBentrySTDinterwordspacing}{\spaceskip=0pt\relax}
\providecommand{\BIBentryALTinterwordstretchfactor}{4}
\providecommand{\BIBentryALTinterwordspacing}{\spaceskip=\fontdimen2\font plus
\BIBentryALTinterwordstretchfactor\fontdimen3\font minus
  \fontdimen4\font\relax}
\providecommand{\BIBforeignlanguage}[2]{{%
\expandafter\ifx\csname l@#1\endcsname\relax
\typeout{** WARNING: IEEEtran.bst: No hyphenation pattern has been}%
\typeout{** loaded for the language `#1'. Using the pattern for}%
\typeout{** the default language instead.}%
\else
\language=\csname l@#1\endcsname
\fi
#2}}
\providecommand{\BIBdecl}{\relax}
\BIBdecl

\bibitem{MiM_Kumar}
J.~T. Chiang, J.~J. Haas, Y.~C. Hu, P.~R. Kumar, and J.~Choi, ``Fundamental
  Limits on Secure Clock Synchronization and Man-In-The-Middle Detection in
  Fixed Wireless Networks,'' in \emph{IEEE INFOCOM 2009}, 2009, pp. 1962--1970.

\bibitem{relay_delay_analysis}
\BIBentryALTinterwordspacing
H.~I. Cho and G.~U. Hwang, ``Packet Delay Analysis of a Wireless Network with
  Multiple Relays under Rayleigh Fading Channels,'' in \emph{Proceedings of the
  5th International Conference on Queueing Theory and Network Applications},
  ser. QTNA '10.\hskip 1em plus 0.5em minus 0.4em\relax New York, NY, USA:
  Association for Computing Machinery, 2010, p. 59–66. [Online]. Available:
  \url{https://doi.org/10.1145/1837856.1837866}
\BIBentrySTDinterwordspacing

\bibitem{relay_capacity}
G.~Kramer, M.~Gastpar, and P.~Gupta, ``Cooperative strategies and capacity
  theorems for relay networks,'' \emph{IEEE Transactions on Information
  Theory}, vol.~51, no.~9, pp. 3037--3063, 2005.

\bibitem{adaptover_mim1}
\BIBentryALTinterwordspacing
S.~Erni, M.~Kotuliak, P.~Leu, M.~Roeschlin, and S.~Capkun, ``AdaptOver:
  Adaptive Overshadowing Attacks in Cellular Networks,'' in \emph{Proceedings
  of the 28th Annual International Conference on Mobile Computing And
  Networking}, ser. MobiCom '22.\hskip 1em plus 0.5em minus 0.4em\relax New
  York, NY, USA: Association for Computing Machinery, 2022, p. 743–755.
  [Online]. Available: \url{https://doi.org/10.1145/3495243.3560525}
\BIBentrySTDinterwordspacing

\bibitem{mitm1}
D.~Rupprecht, K.~Kohls, T.~Holz, and C.~Pöpper, ``Breaking LTE on Layer Two,''
  in \emph{2019 IEEE Symposium on Security and Privacy (SP)}, 2019, pp.
  1121--1136.

\bibitem{4G_sync0}
\BIBentryALTinterwordspacing
{3GPP RAN1}, ``{LTE} Physical layer procedures,'' 2023. [Online]. Available:
  \url{https://www.etsi.org/deliver/etsi_ts/136200_136299/136213/14.02.00_60/ts_136213v140200p.pdf}
\BIBentrySTDinterwordspacing

\bibitem{3gpp_specifications}
\BIBentryALTinterwordspacing
{3GPP working group}, ``\BIBforeignlanguage{en-us}{{Cellular Standards}}.''
  [Online]. Available:
  \url{https://www.3gpp.org/specifications-technologies/specifications-by-series}
\BIBentrySTDinterwordspacing

\bibitem{MiM_Full_Duplex_Implementation}
\BIBentryALTinterwordspacing
S.~Ganji, ``Full Duplex Man-in-the-Middle Implementation using USRP X310,''
  2023. [Online]. Available: \url{https://github.com/shotsan/infinitelooper}
\BIBentrySTDinterwordspacing

\bibitem{wifi1}
H.~Hwang, G.~Jung, K.~Sohn, and S.~Park, ``A Study on MITM (Man in the Middle)
  Vulnerability in Wireless Network Using 802.1X and EAP,'' in \emph{2008
  International Conference on Information Science and Security (ICISS 2008)},
  2008, pp. 164--170.

\bibitem{lynn_baird}
\BIBentryALTinterwordspacing
M.~Lynn and R.~Baird, ``Advanced 802.11 Attack,'' 2002. [Online]. Available:
  \url{http://www.blackhat.com/presentations/bh-usa-02/baird-lynn/bh-us-02-lynn-802.11attack.ppt}
\BIBentrySTDinterwordspacing

\bibitem{cam2003security}
N.~Cam-Winget, R.~Housley, D.~Wagner, and J.~Walker, ``Security flaws in 802.11
  data link protocols,'' \emph{Communications of the ACM}, vol.~46, no.~5, pp.
  35--39, 2003.

\bibitem{bradbury2011hacking}
\BIBentryALTinterwordspacing
D.~Bradbury, ``Feature: Hacking Wifi the Easy Way,'' \emph{Netw. Secur.}, vol.
  2011, no.~2, p. 9–12, feb 2011. [Online]. Available:
  \url{https://doi.org/10.1016/S1353-4858(11)70014-9}
\BIBentrySTDinterwordspacing

\bibitem{ahmad2010wpa}
\BIBentryALTinterwordspacing
M.~S. Ahmad, ``Wpa too!'' 2008, accessed on 09-23-2023. [Online]. Available:
  \url{https://defcon.org/images/defcon-16/dc16-presentations/defcon-16-ahmad.pdf}
\BIBentrySTDinterwordspacing

\bibitem{herzberg2009stealth}
A.~Herzberg and H.~Shulman, ``Stealth-MITM DoS attacks on secure channels,''
  \emph{arXiv preprint arXiv:0910.3511}, 2009.

\bibitem{7031876}
M.~Agarwal, S.~Biswas, and S.~Nandi, ``Advanced Stealth Man-in-The-Middle
  Attack in WPA2 Encrypted Wi-Fi Networks,'' \emph{IEEE Communications
  Letters}, vol.~19, no.~4, pp. 581--584, 2015.

\bibitem{6449834}
V.~Kumar, S.~Chakraborty, F.~A. Barbhuiya, and S.~Nandi, ``Detection of stealth
  Man-in-the-Middle attack in wireless LAN,'' in \emph{2012 2nd IEEE
  International Conference on Parallel, Distributed and Grid Computing}, 2012,
  pp. 290--295.

\bibitem{beam_stealing}
\BIBentryALTinterwordspacing
D.~Steinmetzer, Y.~Yuan, and M.~Hollick, ``Beam-Stealing: Intercepting the
  Sector Sweep to Launch Man-in-the-Middle Attacks on Wireless IEEE 802.11ad
  Networks,'' in \emph{Proceedings of the 11th ACM Conference on Security and
  Privacy in Wireless and Mobile Networks}, ser. WiSec '18.\hskip 1em plus
  0.5em minus 0.4em\relax New York, NY, USA: Association for Computing
  Machinery, 2018, p. 12–22. [Online]. Available:
  \url{https://doi.org/10.1145/3212480.3212499}
\BIBentrySTDinterwordspacing

\bibitem{strobel2007imsi}
D.~Strobel, ``IMSI catcher,'' \emph{Chair for Communication Security,
  Ruhr-Universit{\"a}t Bochum}, vol.~14, 2007.

\bibitem{2G_3G_Mid_generation}
\BIBentryALTinterwordspacing
U.~Meyer and S.~Wetzel, ``A Man-in-the-Middle Attack on UMTS,'' in
  \emph{Proceedings of the 3rd ACM Workshop on Wireless Security}, ser. WiSe
  '04.\hskip 1em plus 0.5em minus 0.4em\relax New York, NY, USA: Association
  for Computing Machinery, 2004, p. 90–97. [Online]. Available:
  \url{https://doi.org/10.1145/1023646.1023662}
\BIBentrySTDinterwordspacing

\bibitem{wifi_calling_tmobile}
J.~G. Beekman and C.~Thompson, ``Breaking Cell Phone Authentication:
  Vulnerabilities in AKA, IMS and Android,'' in \emph{Proceedings of the 7th
  USENIX Conference on Offensive Technologies}, ser. WOOT'13.\hskip 1em plus
  0.5em minus 0.4em\relax USA: USENIX Association, 2013, p.~5.

\bibitem{lte_security_disabled}
\BIBentryALTinterwordspacing
M.~Chlosta, D.~Rupprecht, T.~Holz, and C.~P\"{o}pper, ``LTE Security Disabled:
  Misconfiguration in Commercial Networks,'' in \emph{Proceedings of the 12th
  Conference on Security and Privacy in Wireless and Mobile Networks}, ser.
  WiSec '19.\hskip 1em plus 0.5em minus 0.4em\relax New York, NY, USA:
  Association for Computing Machinery, 2019, p. 261–266. [Online]. Available:
  \url{https://doi.org/10.1145/3317549.3324927}
\BIBentrySTDinterwordspacing

\bibitem{4G_5G}
\BIBentryALTinterwordspacing
A.~Shaik, R.~Borgaonkar, S.~Park, and J.-P. Seifert, ``New Vulnerabilities in
  4G and 5G Cellular Access Network Protocols: Exposing Device Capabilities,''
  in \emph{Proceedings of the 12th Conference on Security and Privacy in
  Wireless and Mobile Networks}, ser. WiSec '19.\hskip 1em plus 0.5em minus
  0.4em\relax New York, NY, USA: Association for Computing Machinery, 2019, p.
  221–231. [Online]. Available: \url{https://doi.org/10.1145/3317549.3319728}
\BIBentrySTDinterwordspacing

\bibitem{Sigunder}
\BIBentryALTinterwordspacing
N.~Ludant and G.~Noubir, ``SigUnder: A Stealthy 5G Low Power Attack and
  Defenses,'' in \emph{Proceedings of the 14th ACM Conference on Security and
  Privacy in Wireless and Mobile Networks}, ser. WiSec '21.\hskip 1em plus
  0.5em minus 0.4em\relax New York, NY, USA: Association for Computing
  Machinery, 2021, p. 250–260. [Online]. Available:
  \url{https://doi.org/10.1145/3448300.3467817}
\BIBentrySTDinterwordspacing

\bibitem{Rupprecht2020IMP4GTIA}
D.~Rupprecht, K.~S. Kohls, T.~Holz, and C.~P{\"o}pper, ``IMP4GT: IMPersonation
  Attacks in 4G NeTworks,'' in \emph{Network and Distributed System Security
  Symposium}, 2020.

\bibitem{RSS}
T.~Kim, H.~Park, H.~Jung, and H.~Lee, ``Online Detection of Fake Access Points
  Using Received Signal Strengths,'' in \emph{2012 IEEE 75th Vehicular
  Technology Conference (VTC Spring)}, 2012, pp. 1--5.

\bibitem{TCP_ACKs}
\BIBentryALTinterwordspacing
W.~Wei, K.~Suh, B.~Wang, Y.~Gu, J.~Kurose, and D.~Towsley, ``Passive Online
  Rogue Access Point Detection Using Sequential Hypothesis Testing with TCP
  ACK-Pairs,'' in \emph{Proceedings of the 7th ACM SIGCOMM Conference on
  Internet Measurement}, ser. IMC '07.\hskip 1em plus 0.5em minus 0.4em\relax
  New York, NY, USA: Association for Computing Machinery, 2007, p. 365–378.
  [Online]. Available: \url{https://doi.org/10.1145/1298306.1298357}
\BIBentrySTDinterwordspacing

\bibitem{IPAT}
R.~Beyah, S.~Kangude, G.~Yu, B.~Strickland, and J.~Copeland, ``Rogue access
  point detection using temporal traffic characteristics,'' in \emph{IEEE
  Global Telecommunications Conference, 2004. GLOBECOM '04.}, vol.~4, 2004, pp.
  2271--2275 Vol.4.

\bibitem{RTT}
H.~Han, B.~Sheng, C.~C. Tan, Q.~Li, and S.~Lu, ``A Timing-Based Scheme for
  Rogue AP Detection,'' \emph{IEEE Transactions on Parallel and Distributed
  Systems}, vol.~22, no.~11, pp. 1912--1925, 2011.

\bibitem{RTT1}
G.~Qu and M.~N. Michael, ``RAPiD: An indirect rogue access points detection
  system,'' in \emph{International Performance Computing and Communications
  Conference}, 2010, pp. 9--16.

\bibitem{traffic}
R.~Beyah, S.~Kangude, G.~Yu, B.~Strickland, and J.~Copeland, ``Rogue access
  point detection using temporal traffic characteristics,'' in \emph{IEEE
  Global Telecommunications Conference, 2004. GLOBECOM '04.}, vol.~4, 2004, pp.
  2271--2275 Vol.4.

\bibitem{ieee_standards}
\BIBentryALTinterwordspacing
IEEE, ``{IEEE} Wired and Wireless {Standards}.'' [Online]. Available:
  \url{https://ieeexplore.ieee.org/browse/standards/get-program/page/series?id=68}
\BIBentrySTDinterwordspacing

\bibitem{bluetooth}
\BIBentryALTinterwordspacing
{Bluetooth Special Interest Group}, ``{Bluetooth}® {Technology}
  Specifications.'' [Online]. Available:
  \url{https://www.bluetooth.com/specifications/specs/}
\BIBentrySTDinterwordspacing

\bibitem{5G_sync}
\BIBentryALTinterwordspacing
3GPP, ``NewRadio physical layer procedures for control,'' 2023. [Online].
  Available:
  \url{https://www.etsi.org/deliver/etsi_ts/138200_138299/138213/15.05.00_60/ts_138213v150500p.pdf}
\BIBentrySTDinterwordspacing

\bibitem{sullivan1990characterization}
D.~B. Sullivan, D.~W. Allan, D.~A. Howe, D.~Sullivan, and F.~Walls,
  ``Characterization of clocks and oscillators,'' 1990.

\bibitem{impossitility_theorem_proof}
N.~M. Freris, S.~R. Graham, and P.~R. Kumar, ``Fundamental Limits on
  Synchronizing Clocks Over Networks,'' \emph{IEEE Transactions on Automatic
  Control}, vol.~56, no.~6, pp. 1352--1364, 2011.

\bibitem{PTP}
J.~C. Eidson, M.~Fischer, and J.~White, ``IEEE-1588™ Standard for a precision
  clock synchronization protocol for networked measurement and control
  systems,'' in \emph{Proceedings of the 34th Annual Precise Time and Time
  Interval Systems and Applications Meeting}, 2002, pp. 243--254.

\bibitem{Mills_ptp_ntp}
\BIBentryALTinterwordspacing
D.~L. Mills, ``Details on NTP and PTP,'' accessed on 09-23-2023. [Online].
  Available: \url{https://www.eecis.udel.edu/~mills/ptp.html}
\BIBentrySTDinterwordspacing

\bibitem{5gcqi}
A.~K. Thyagarajan, P.~Balasubramanian, V.~D, and K.~M, ``SNR-CQI Mapping for 5G
  Downlink Network,'' in \emph{2021 IEEE Asia Pacific Conference on Wireless
  and Mobile (APWiMob)}, 2021, pp. 173--177.

\bibitem{ghosh2011essentials}
A.~Ghosh and R.~Ratasuk, \emph{Essentials Of Lte And Lte-A}.\hskip 1em plus
  0.5em minus 0.4em\relax Cambridge University Press, 2011.

\bibitem{srsran}
\BIBentryALTinterwordspacing
srsRAN Project, ``open source 4G software radio suite,'' Jan 2023. [Online].
  Available: \url{https://github.com/srsran/srsRAN_4G}
\BIBentrySTDinterwordspacing

\bibitem{usrpx310}
\BIBentryALTinterwordspacing
{Ettus Research}, ``USRP X310,'' Jan 2023. [Online]. Available:
  \url{https://www.ettus.com/all-products/X310-KIT/}
\BIBentrySTDinterwordspacing

\bibitem{usrpb210}
\BIBentryALTinterwordspacing
{Ettus Research}., ``USRP B210,'' Jan 2023. [Online]. Available:
  \url{https://www.ettus.com/all-products/ub210-kit/}
\BIBentrySTDinterwordspacing

\bibitem{OAI-5G}
\BIBentryALTinterwordspacing
{Open Air 5G Team }, ``Open air 5G project,'' accessed on 10/24/2023. [Online].
  Available: \url{https://openairinterface.org/oai-5g-ran-project/}
\BIBentrySTDinterwordspacing

\bibitem{srsran5G}
\BIBentryALTinterwordspacing
srsRAN Project, ``5G O-RAN CU/DU solution,'' accessed on 10/24/2023. [Online].
  Available: \url{https://www.srsran.com/5g}
\BIBentrySTDinterwordspacing

\bibitem{MIM_UE}
\BIBentryALTinterwordspacing
S.~Ganji, ``A slightly modified srsRAN 4G eNodeB and UE for better debugging,''
  2023. [Online]. Available: \url{https://github.com/shotsan/srsue_mim}
\BIBentrySTDinterwordspacing

\bibitem{Time_sync_scripts1}
\BIBentryALTinterwordspacing
S.~Ganji., ``Clock synchronization scripts,'' 2023. [Online]. Available:
  \url{https://github.com/shotsan/mim_server_client}
\BIBentrySTDinterwordspacing

\bibitem{mim_demo}
\BIBentryALTinterwordspacing
S.~Ganji and P.~R. Kumar, ``Seeing the Unseen: The REVEAL Protocol to Catch the
  Wireless Man-in-the-Middle,'' in \emph{Proceedings of the 14th ACM Wireless
  of the Students, by the Students, and for the Students (S3) Workshop}, ser.
  S3 '23.\hskip 1em plus 0.5em minus 0.4em\relax New York, NY, USA: Association
  for Computing Machinery, 2023, p. 3–4. [Online]. Available:
  \url{https://doi.org/10.1145/3615591.3615678}
\BIBentrySTDinterwordspacing

\bibitem{demos}
\BIBentryALTinterwordspacing
{Santosh Ganji}., ``Demonstration videos of the working system,'' Jan 2023.
  [Online]. Available:
  \url{https://github.com/shotsan/MiM_paper_demo_videos/tree/main}
\BIBentrySTDinterwordspacing

\bibitem{full_duplex_arc}
\BIBentryALTinterwordspacing
D.~Bharadia, E.~McMilin, and S.~Katti, ``Full Duplex Radios,'' \emph{SIGCOMM
  Comput. Commun. Rev.}, vol.~43, no.~4, p. 375–386, aug 2013. [Online].
  Available: \url{https://doi.org/10.1145/2534169.2486033}
\BIBentrySTDinterwordspacing

\bibitem{cell_towerdatabase}
\BIBentryALTinterwordspacing
{Open Cell-id Organisation.}, ``The world's largest Open Database of Cell
  Towers,'' Jan 2023. [Online]. Available:
  \url{https://www.opencellid.org/#zoom=16&lat=37.77889&lon=-122.41942}
\BIBentrySTDinterwordspacing

\end{thebibliography}
\end{document}